\documentclass[aps,twocolumn,groupedaddress,showpacs,notitlepage,nofootinbib]{revtex4}
\usepackage{graphicx}
\usepackage{dcolumn}
\usepackage{bm}
\usepackage{amsmath}
\usepackage{amsthm}
\usepackage{multirow}
\usepackage{enumerate}
\usepackage{amsfonts}
\usepackage{ifthen}
\usepackage{psfrag}
\usepackage{slashed}
\usepackage{gensymb}
\usepackage[utf8]{inputenc}
\usepackage{color}
\usepackage{subfigure}
\usepackage{amssymb}

\usepackage[colorlinks, citecolor=blue, linkcolor=red, urlcolor=blue]{hyperref}
\begin{document}
\newcommand{\psl}{ p \hspace{-1.8truemm}/ }
\newcommand{\nsl}{ n \hspace{-2.2truemm}/ }
\newcommand{\vsl}{ v \hspace{-2.2truemm}/ }
\newcommand{\epsl}{\epsilon \hspace{-1.8truemm}/\,}
\renewcommand{\arraystretch}{1.5}
\title{Quasi-two-body decays $B^+\to D_s^+ (R\to) K^+K^-$ in the perturbative QCD approach}
\author{Zhi-Tian Zou}\email{zouzt@ytu.edu.cn}
\author{Jun-Peng Wang}
\author{Zhou Rui}\email{jindui1127@126.com}
\author{Ying Li}\email{liying@ytu.edu.cn}

\affiliation{Department of Physics, Yantai University, Yantai 264005, China}
\date{\today}
\begin{abstract}
A search for the decay $B^+\to D_s^+ K^+K^-$ has been reported by the LHCb Collaboration using $pp$ collision data corresponding to an integrated luminosity of $4.8\,\mathrm{fb}^{-1}$, collected at center-of-mass energies of 7, 8, and 13~TeV, in which no amplitude analysis of the $K^+K^-$ subsystem was performed. In this work, we study the resonant contributions to the decay $B^+\to D_s^+ K^+K^-$ within the perturbative QCD (PQCD) factorization framework. Contributions from the $S$-wave resonances $f_0(980)$, $f_0(1370)$, and $f_0(1500)$, the $P$-wave resonance $\phi(1020)$, and the $D$-wave resonances $f_2(1270)$ and $f_2(1525)$ are taken into account. By introducing the corresponding two-meson distribution amplitudes for the $K^+K^-$ system, we perform a complete perturbative analysis of the quasi-two-body decays $B^+\to D_s^+(R\to)K^+K^-$, where $R$ denotes an intermediate resonance, and present the first PQCD predictions for the associated branching fractions. Using the narrow-width approximation, we further extract the branching fractions of the corresponding two-body decays $B^+\to D_s^+R$. Our results are consistent with the available experimental measurements and previous theoretical studies. Finally, we find that direct CP asymmetries vanish for these quasi-two-body decays within the Standard Model, so that any experimentally observed nonzero CP asymmetry would constitute a clear signal of physics beyond the Standard Model.

\end{abstract}
\maketitle

\section{Introduction}
It is well established that $B$-meson decays provide an excellent laboratory for precision tests of the Standard Model (SM), investigations of the origin of CP violation, and searches for physics beyond the SM. Compared with two-body decays, three-body $B$-meson decay modes are significantly more abundant and exhibit a much richer phenomenology. Their decay dynamics are intrinsically more involved, featuring nontrivial interference among multiple contributing amplitudes, which naturally generates strong phases and enhances sensitivity to CP-violating effects. As an illustrative example, the decays $B_s^0\to K_S^0 K^{\pm}\pi^{\mp}$ are not flavor specific: both $B_s^0$ and $\bar B_s^0$ mesons can decay into the final states $K_S^0 K^- \pi^+$ and $K_S^0 K^+ \pi^-$ with amplitudes of comparable magnitude. The sizable interference between these contributions can therefore induce large CP asymmetries, providing a particularly promising environment for CP violation studies. Over the past two decades, extensive experimental investigations of three-body $B$-meson decays have been carried out by the BaBar~\cite{BaBar:2009vfr, BaBar:2008lpx, BaBar:2011ktx, BaBar:2012iuj}, Belle~\cite{Belle:2004drb, Belle:2005rpz, Belle:2010wis, Belle:2006ljg}, and LHCb~\cite{LHCb:2017hbp, LHCb:2019sus, LHCb:2022oqs, LHCb:2025qcs, LHCb:2024vhs, LHCb:2022lzp, LHCb:2019vww, LHCb:2022dvn, LHCb:2024vfz, LHCb:2023qca} collaborations, yielding high-precision measurements of branching fractions and CP-violating observables. These results provide stringent tests of the SM and valuable opportunities to further elucidate the dynamics of CP violation and to probe possible new-physics effects.

In two-body decays, the kinematics are fixed, and the heavy-quark limit can be exploited in a systematic manner within QCD factorization~\cite{Chen:1999nxa,Cheng:2008gxa,Beneke:2003zv}, soft-collinear effective theory~\cite{Bauer:2000yr,Beneke:2002ph}, and the perturbative QCD (PQCD) approach~\cite{Lu:2000em,Lu:2000hj,Ali:2007ff,Li:1994cka,Li:1995jr}, where the hadronic matrix elements factorize into convolutions of perturbatively calculable hard kernels with nonperturbative inputs such as form factors and light-cone distribution amplitudes of the participating mesons. In this framework, long-distance final-state interactions are parametrically suppressed in the heavy-quark limit, aided by the fixed two-body kinematics and the large energy release. In contrast, three-body decay amplitudes depend on two independent kinematic variables and receive contributions from both resonant and nonresonant components, as well as potentially sizable final-state interactions among the three outgoing mesons. Many three-body $B$-meson decays are dominated by intermediate vector or scalar resonances and proceed predominantly through quasi-two-body channels involving a resonant subsystem accompanied by a bachelor meson, whereas in certain modes nonresonant three-body contributions are found to be substantial. Based on comprehensive experimental amplitude analyses, it has been established that nonresonant contributions play a dominant role in penguin-dominated three-body $B$-meson decays. For example, in $B\to KKK$ decays the nonresonant fraction can reach approximately $70\%$--$90\%$, while in the $B\to K\pi\pi$ and $B\to \pi\pi\pi$ channels the corresponding fractions are smaller, at about $40\%$ and $14\%$, respectively \cite{Cheng:2008vy}.

A clear separation between resonant and nonresonant contributions is therefore essential for a reliable theoretical description of three-body $B$-meson decays. The Dalitz plot provides a natural framework for this purpose, allowing the phase space to be organized into regions characterized by distinct kinematic configurations. The central region corresponds to configurations in which all three final-state particles carry comparable energies, $E\sim M_B/3$, in the $B$-meson rest frame and none of the particles are collinear. The edge regions are associated with kinematics in which two of the final-state mesons move collinearly, forming an energetic subsystem with an invariant mass that recoils against a third bachelor meson. The corner regions describe configurations in which one of the final-state mesons is soft or nearly at rest, while the remaining two mesons move back-to-back with large energies of order $E\sim M_B/2$. In this work, we concentrate on the dynamics near the edges of the Dalitz plot, where the two nearly collinear mesons can be treated as a clustered system. Within this cluster, the two mesons may form intermediate resonant states with different quantum numbers, leading to quasi-two-body decay topologies beyond the narrow-width approximation. From the theoretical perspective, three-body $B$-meson decays have been extensively investigated using a variety of approaches, including QCD factorization~\cite{El-Bennich:2009gqk,Krankl:2015fha,Cheng:2002qu,Cheng:2016shb,Li:2014fla,Li:2014oca}, the perturbative QCD approach~\cite{Wang:2014ira,Li:2016tpn,Rui:2017bgg,Zou:2020atb,Zou:2020fax,Zou:2020mul,Zou:2020ool,Yang:2021zcx}, as well as other theoretical frameworks~\cite{Hu:2022eql,Zhang:2013oqa,Cheng:2019tgh}.

To date, in addition to the numerous charmless three-body $B$-meson decay modes measured at the $B$ factories and by the LHCb Collaboration, an increasing number of charmed three-body decay modes containing one or two $D_{(s)}$ mesons in the final state have been studied in detail, both at the $B$ factories~\cite{BaBar:2009pnd} and at LHCb~\cite{LHCb:2013svv,LHCb:2014ioa,LHCb:2015tsv,LHCb:2016bsl,LHCb:2018oeb}. Compared with charmless $B$ decays, open-charm $B$-meson decays are dominated by tree-level amplitudes in SM, rendering them theoretically cleaner for the determination of the Cabibbo--Kobayashi--Maskawa (CKM) unitarity-triangle angle $\gamma$ and sensitive probes of possible new-physics effects~\cite{Charles:2013aka}. For instance, the decay $B_s^0\to \bar D^0\phi$ has the potential to significantly improve the precision on $\gamma$, while a Dalitz-plot analysis of the related three-body decay $B_s^0\to \bar D^0 K^+K^-$ can further enhance the sensitivity by exploiting interference effects among different intermediate contributions~\cite{LHCb:2018oeb}. In contrast to charmless three-body $B$ decays, charmed three-body $B$-meson decays exhibit a richer spectrum of intermediate resonant states: in addition to resonances formed by two light mesons, resonant structures composed of a $D$ meson and a light meson can also be accessed, such as the tetraquark candidate $T^*_{cs0}(2870)^0$ observed in the $D^+K^-$ system~\cite{LHCb:2024xyx}. Consequently, charmed three-body $B$-meson decays are expected to display a rich dynamical structure with multiple intermediate resonances, and their detailed analyses provide valuable information on resonance properties, polarization fractions, and potential signals of physics beyond the SM. On the theoretical side, quasi-two-body decays of the type $B_{(s)}\to [D^{(*)},\bar D^{(*)}]K^+K^-$ were systematically investigated within the PQCD framework in Ref.~\cite{Li:2020zng}, while the resonant contributions to $B\to KKK$ decays, including $S$-, $P$-, and $D$-wave $KK$ resonances, were studied in Ref.~\cite{Zou:2020atb}. In the present work, we extend these studies to the quasi-two-body decays $B^+\to D_s^+(R\to)K^+K^-$ by incorporating the contributions from $S$-, $P$-, and $D$-wave $K^+K^-$ resonances, while resonant effects associated with the $D_s^+K^-$ subsystem are not considered and are deferred to future studies due to their additional dynamical complexity.

In quasi-two-body decays occurring near the edges of the Dalitz plot, one invariant mass becomes small, and low-energy interactions between the corresponding pair of final-state mesons give rise to resonant structures, while the interaction between this meson pair and the bachelor meson is strongly suppressed. As a result, such decay topologies effectively resemble two-body decays, with the essential difference that one of the final-state particles is replaced by a correlated two-meson system. By introducing appropriate two-meson wave functions, the factorization formalism for quasi-two-body decays attains a level of theoretical rigor comparable to that of genuine two-body decays, even beyond the narrow-width approximation, which constitutes a key advantage of this approach. Within the perturbative QCD (PQCD) framework, the amplitude for a quasi-two-body nonleptonic $B$-meson decay can be systematically factorized according to the relevant energy scales: physics above the $W$-boson mass $m_W$ is governed by weak interactions and treated perturbatively; effects between the scales $m_W$ and the $b$-quark mass $m_b$ are encoded in the Wilson coefficients $C(\mu)$, evolved via renormalization-group equations from their perturbative matching at $m_W$; contributions between $m_b$ and the factorization scale $\Lambda_h$ arise from hard gluon exchange and are described by a channel-dependent hard kernel $H$, calculable perturbatively; finally, dynamics below $\Lambda_h$ are soft and nonperturbative, and are parameterized by universal wave functions of the participating hadrons. Accordingly, the decay amplitude for the quasi-two-body process $B^{+}\to D_s^{+}(R\to)K^{+}K^{-}$ can be expressed in the convolution form~\cite{Li:2003yj}
\begin{multline}
    \mathcal{A} \sim\int dx_idb_i\Big[ C(t)\otimes H(x_i,b_i,t)\otimes \Phi_B(x_1,b_1) 
    \\\otimes \Phi_{KK}(x_2,b_2)\otimes \Phi_D(x_3,b_3)\otimes e^{-S(t)}\Big] ,
    \label{eq:amplitude}
\end{multline}
where $x_i$ denote the longitudinal momentum fractions carried by the quarks in the initial- and final-state hadrons, $b_i$ are the conjugate variables to the transverse momenta $k_{Ti}$, $C(t)$ are the Wilson coefficients, $\Phi_B$ and $\Phi_D$ represent the wave functions of the $B$ and $D$ mesons, respectively, and $\Phi_{KK}$ is the two-meson wave function of the $K^+K^-$ system. The hard kernel $H$ is perturbatively calculable within PQCD, while the Sudakov factor $e^{-S(t)}$, originating from threshold resummation, suppresses soft contributions and removes endpoint divergences.

The organization of this paper is as follows. In Sec.~\ref{sec:2}, we present the theoretical framework for the quasi-two-body decays $B^+\to D_s^+(R\to)K^+K^-$ and introduce the relevant nonperturbative inputs, in particular the two-meson wave functions describing the $K^+K^-$ system. In Sec.~\ref{sec:3}, we derive the analytical expressions for the decay amplitudes and present the corresponding numerical results for the quasi-two-body decay channels under consideration. Employing the narrow-width approximation, we further evaluate the branching fractions of the related two-body decays $B^+\to D_s^+R$, where $R$ denotes the intermediate resonances decaying into the $K^+K^-$ pair. Finally, a brief summary and concluding remarks are given in the last section.

\section{Framework} \label{sec:2}
The theoretical description of three-body $B$-meson decays is well recognized to be in the modeling phase. A key challenge is the model-dependent parametrization of resonant structures, as the line shapes of strong resonances vary across different theoretical approaches. Dalitz plot analysis plays a key role in experimental amplitude analyses of multi-body decays. Using this technique, the isobar model allows the decomposition of the total amplitude for quasi-two-body decays into a coherent sum of contributions from $N$ distinct resonant channels, each associated with a specific intermediate resonance. The total amplitude can thus be written as
\begin{equation}
    \mathcal{A} = \sum a_i \mathcal{A}_i
    \label{ibm}
\end{equation}
where $a_i$ is the complex coefficient that encodes the relative magnitude and strong phase of the $i$-th channel. The relative strong phase between different resonant amplitudes provides a novel mechanism for $CP$ violation, arising from quantum interference between distinct resonant pathways. The partial amplitude $\mathcal{A}_i$ for each resonant channel can be computed perturbatively within the PQCD framework, with the general expression given in Eq.~\eqref{eq:amplitude}.

\begin{figure*}[!htb]
    \centering
    \includegraphics[width=1\linewidth]{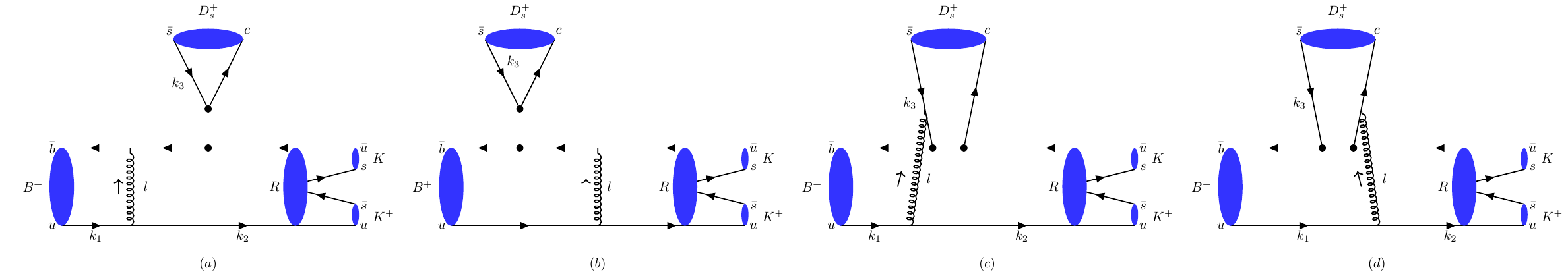}
    \caption{ Leading order emission type Feynman diagrams for the $B^+\to D_{s}^{+}( f_{0} /f_{2}\to ) K^{+} K^{-}$ decays. }
    \label{fig:placeholder1}
\end{figure*}

\begin{figure*}[!htb]
    \centering
    \includegraphics[width=1\linewidth]{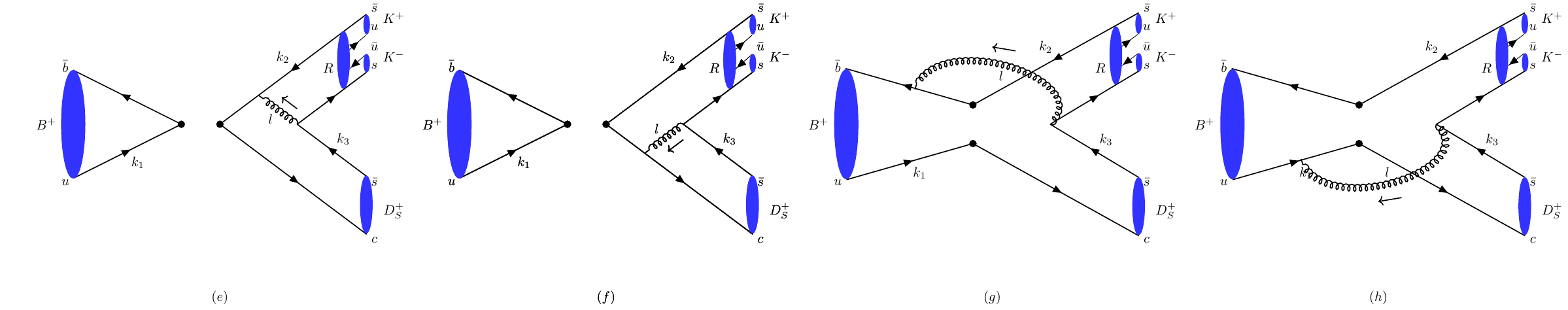}
    \caption{Leading order annihilation type Feynman diagrams contributing to $B^+\to D_{s}^{+}(R\to ) K^{+} K^{-}$ }
    \label{fig:placeholder2}
\end{figure*}

To simplify the treatment of momentum variables, we work in the rest frame of the $B$ meson and adopt the light-cone coordinate system. The $K^+K^-$ pair and the $D_s^+$ meson are assumed to move along the $n = (1, 0, \mathbf{0}_T)$ and $v = (0, 1, \mathbf{0}_T)$ directions, respectively. The momenta of the $B^+$ meson, the total momentum of the $K^+K^-$ pair, and the $D_s^+$ meson are denoted by $P_1$, $P_2$, and $P_3$, respectively, while $k_i$ represents the valence quark momentum in both the initial and final states (depicted in Figs.~\ref{fig:placeholder1} and \ref{fig:placeholder2}). These momenta are given by:
\begin{align}
P_1 &= \frac{m_B}{\sqrt{2}} (1, 1, \mathbf{0}_T),  \nonumber \\
P_2 &= \frac{m_B}{\sqrt{2}} \left( 1 - r_d^2, \; \eta^2, \; \mathbf{0}_T \right),  \nonumber \\
P_3 &= \frac{m_B}{\sqrt{2}} \left( r_d^2, \; 1 - \eta^2, \; \mathbf{0}_T \right),
\end{align}
and
\begin{align}
k_1 &= \left( 0, \; \frac{m_B}{\sqrt{2}} x_1, \; \mathbf{k}_{1T} \right), \nonumber \\
k_2 &= \left( (1 - r_d^2) \frac{m_B}{\sqrt{2}} x_2, \; 0, \; \mathbf{k}_{2T} \right), \nonumber \\
k_3 &= \left( 0, \; \frac{m_B}{\sqrt{2}} (1 - \eta^2) x_3, \; \mathbf{k}_{3T} \right),
\end{align}
with
\begin{equation}
   \eta^2 = \frac{\omega^2}{m_B^2(1 - r_d^2)}
    \label{eta}
\end{equation}
Here, $m_B$ is the mass of the $B^+$ meson and $r_d = {m_D}/{m_B}$. The invariant mass of the $K^+K^-$ pair is defined as $\omega^2 = (p_1 + p_2)^2 = P_2^2$, where $p_{1,2}$ are the momenta of the two kaons. The variables $x_1$, $x_2$, and $x_3$ represent the momentum fractions of the light quark in the $B^+$, $D_s^+$, and the resonant structure, respectively, with each ranging from 0 to 1. In the heavy quark limit, we can ignore the mass differences between the $b$-quark and the $B$-meson, as well as between the $c$-quark and the $D$-meson.


To calculate the decay amplitude, the first step is to determine the effective Hamiltonian governing the weak interaction. In the SN, the effective Hamiltonian $\mathcal{H}_{\text{eff}}$ for the quasi-two-body decay $B^{+} \to D_s^{+} (R \to) K^{+} K^{-}$ is given by \cite{Shen:2014wga}
\begin{align}
    \mathcal{H}_{\text{eff}} = \frac{G_F}{\sqrt{2}} V_{ub}^* V_{cs} \left[ C_1(\mu) O_1(\mu) + C_2(\mu) O_2(\mu) \right],
    \label{Hmtn}
\end{align}
where $G_F$ is the Fermi constant, $V_{ub}$ and $V_{cs}$ are the Cabibbo-Kobayashi-Maskawa (CKM) matrix elements, $O_1$ and $O_2$ are the tree-level local four-quark operators, and $C_1, C_2$ are their corresponding Wilson coefficients at the renormalization scale $\mu$. The tree-level operators $O_{1,2}$ are defined as
\begin{align}
    O_1 &= \bar{b}_\alpha \gamma^\mu (1 - \gamma_5) u_\beta \, \bar{c}_\beta \gamma_\mu (1 - \gamma_5) s_\alpha, \nonumber \\
    O_2 &= \bar{b}_\alpha \gamma^\mu (1 - \gamma_5) u_\alpha \, \bar{c}_\beta \gamma_\mu (1 - \gamma_5) s_\beta.
\end{align}
It is important to note that the direct $CP$ asymmetry vanishes in these quasi-two-body decays, as it depends on the interference between contributions from tree operators and penguin operators. However, the Hamiltonian contains no penguin operators, preventing such interference.

From Eq.~\eqref{eq:amplitude}, it is obvious that the wave functions of both the initial and final states are critical to calculating the decay amplitude. In the PQCD approach, the wave function serves as the primary nonperturbative input. The wave functions for the initial $B^+$ meson and the final $D_s^+$ meson have been extensively studied in numerous two-body decays, including $B \to PP, PV, VV$ \cite{Ali:2007ff, Zou:2015iwa}, and $B \to D_sP, D_sV, D_sS, D_sT$ \cite{Li:2008ts, Zou:2012sx, Zou:2012sy, Zou:2016yhb, Zou:2017iau, Zou:2017yxc}, with these wave functions already well-determined. For brevity, we will not revisit these wave functions in this work. 

The two-meson wave function, which serves as a new nonperturbative input in quasi-two-body decays, varies with the intermediate resonances. This will be discussed in detail below. In this study, we focus on the resonant contributions of the scalar resonances $f_0(980)$, $f_0(1370)$, and $f_0(1500)$, the vector resonance $\phi(1020)$, and the tensor resonances $f_2(1270)$ and $f^\prime_2(1525)$ to the quasi-two-body decay $B^{+} \to D_s^{+} (R \to) K^{+} K^{-}$. It should be noted that a precise two-meson wave function derived from a QCD-inspired approach has not yet been fully established and remains at the modeling stage. In this work, we adopt a phenomenological model where the nonperturbative parameters are determined by fitting to available experimental data.

We first introduce the $S$-wave wave function for the $K-K$ pair, which involves scalar intermediate resonances such as $f_0(980)$, $f_0(1370)$, and $f_0(1500)$. The $S$-wave wave function for the $K-K$ pair has been carefully determined in Refs.~\cite{Zou:2020atb, Wang:2025cjt, Li:2020zng}, where the explicit form for $\Phi_{S,KK}$ is expressed as
\begin{widetext}
\begin{align}
    \Phi_{S,KK} = \frac{1}{2\sqrt{N_c}} \left[ \slashed{p}_2 \phi_S^0(z, \zeta, \omega^2) + \omega \phi_S^s(z, \zeta, \omega^2) + \omega (\slashed{n} \slashed{v} - 1) \phi_S^t(z, \zeta, \omega^2) \right],
\end{align}
\end{widetext}
where $P_2$ and $\omega$ represent the momentum and invariant mass of the $K-K$ pair, respectively, with $P_2^2 = \omega^2$ following the relativistic energy-momentum relation. The functions $\phi_S^0$, $\phi_S^s$, and $\phi_S^t$ are the twist-2 and twist-3 light-cone distribution amplitudes (LCDAs), respectively. For the twist-3 LCDAs, which are suppressed relative to the twist-2 LCDA, we adopt the asymptotic forms for simplicity and precision. The parameter $z$ denotes the momentum fraction carried by the spectator quark, while $\zeta$ represents the momentum fraction of one $K$ meson in the $K-K$ pair. These LCDAs are typically expanded in terms of Gegenbauer polynomials, which are characterized by corresponding Gegenbauer moments. The explicit forms of the LCDAs are as follows:
\begin{widetext}
\begin{align}
    \phi_S^0(z, \zeta, \omega^2) &= \frac{6F_S(\omega^2)}{2\sqrt{2N_c}} z(1-z) \left[a_{1S} C_1^{3/2}(1-2z) + a_{3S} C_3^{3/2}(1-2z) \right], \\
    \phi_S^s(z, \zeta, \omega^2) &= \frac{F_S(\omega^2)}{2\sqrt{2N_c}}, \\
    \phi_S^t(z, \zeta, \omega^2) &= \frac{F_S(\omega^2)}{2\sqrt{2N_c}}(1-2z),
\end{align}
\end{widetext}
where $C_n^{3/2}(n=1,3)$ are the Gegenbauer polynomials, and $a_{1S}$, $a_{3S}$ are the corresponding Gegenbauer moments. These polynomials, $C_1^{3/2}$ and $C_3^{3/2}$, have standard forms available in tables of special functions. However, the values of $a_{1S}$ and $a_{3S}$ have not been determined by QCD-inspired methods and are currently probed via phenomenological approaches. In Ref.~\cite{Zou:2020atb}, we investigated the $B \to KKK$ decays, where we considered only the leading Gegenbauer polynomial $C_1^{3/2}$ and neglected the mass of the $K$ meson. We obtained the value of the Gegenbauer moment $a_{1S} = -0.8$. In Ref.~\cite{Wang:2025cjt}, the analysis was extended to the decays $B \to \pi/K(f_0(980) \to) K^+K^-$, where the masses of the light final-state mesons were retained, and more precise two-meson wave functions were adopted. The $C_3^{3/2}$ term was also taken into account. By fitting the theoretical results to current experimental data, we determined the values $a_{1S} = -0.8$ and $a_{3S} = 0.2$. These values, as reported in Ref.~\cite{Wang:2025cjt}, are adopted in the present work. It is important to note that the $\zeta$-dependent term in the LCDAs vanishes because the $\zeta$-dependent function is a Legendre polynomial, $P_0(2\zeta-1)$, which equals 1 for all $S$-wave LCDAs.

A key difference between the two-meson wave function and the meson wave function lies in the time-like form factor $F(\omega^2)$. As previously mentioned, this form factor lacks a universal description and remains in the modeling stage. For narrow-width resonances, the relativistic Breit-Wigner (RBW) model \cite{ParticleDataGroup:2022pth} effectively describes the time-like form factor and is widely employed in experimental analyses. A detailed discussion of the time-like form factor will follow.

For the $P$-wave $K-K$ pair, corresponding to a vector resonant structure, the $P$-wave two-meson wave function is available in Refs.~\cite{Zou:2020atb, Li:2020zng}. Due to angular momentum conservation, only the longitudinal polarization component contributes to the quasi-two-body decay $B^+ \to D_s^+ (K^+ K^-)_P$. The longitudinal wave function is given by
\begin{widetext}
\begin{align}
    \Phi_P = \frac{1}{\sqrt{2N_c}} \left[ \slashed{P}_2 \phi_P^0(z, \zeta, \omega^2) + \omega \phi_P^s(z, \zeta, \omega^2) + \frac{\slashed{p}_1 \slashed{p}_2 - \slashed{p}_2 \slashed{p}_1}{\omega (2\zeta - 1)} \phi_P^t(z, \zeta, \omega^2) \right],
    \label{eq:pwave}
\end{align}
\end{widetext}
where $\phi_P^0$ is the twist-2 LCDA, and $\phi_P^s$ and $\phi_P^t$ are the twist-3 LCDAs. These can be expanded in terms of Gegenbauer polynomials as follows:
\begin{widetext}
\begin{align}
    \phi_P^0(z, \zeta, \omega^2) &= \frac{3F_P^{\parallel}(\omega^2)}{\sqrt{2N_c}} z(1-z) \left[1 + a_2^{0P} \frac{3}{2} \left(5(1-2z)^2 - 1 \right) \right] P_1(2\zeta - 1), \\
    \phi_P^s(z, \zeta, \omega^2) &= \frac{3F_P^{\perp}(\omega^2)}{2\sqrt{2N_c}} (1 - 2z) \left[1 + a_2^{sP} (10z^2 - 10z + 1) \right] P_1(2\zeta - 1), \\
    \phi_P^t(z, \zeta, \omega^2) &= \frac{3F_P^{\perp}(\omega^2)}{2\sqrt{2N_c}} (1 - 2z)^2 \left[1 + a_2^{tP} \frac{3}{2} \left( 5(1 - 2z)^2 - 1 \right) \right] P_1(2\zeta - 1).
\end{align}
\end{widetext}
Here, $F_P^{\parallel}(\omega^2)$ and $F_P^{\perp}(\omega^2)$ are the $P$-wave time-like form factors, which will be introduced later. The Gegenbauer moments $a_2^{0P}$, $a_2^{sP}$, and $a_2^{tP}$ are determined as $-0.89 \pm 0.18$, $-0.87 \pm 0.18$, and $0.1 \pm 0.02$, respectively. In Ref.~\cite{Zou:2020atb}, we adopted the same model for the $P$-wave two-meson wave function, neglecting the mass of the $K$ meson, and obtained the values of the Gegenbauer moments as $-0.6 \pm 0.12$, $-0.8 \pm 0.16$, and $-0.3 \pm 0.06$, respectively, from fits to experimental data on $B \to KKK$ decays. In Ref.~\cite{Li:2020zng}, the Gegenbauer moments were determined as $-0.50 \pm 0.10$, $-0.70 \pm 0.14$, and $-0.30 \pm 0.06$, respectively. We note that for $a_2^{0P}$ and $a_2^{sP}$, the values are in agreement within the uncertainties, while the value of $a_2^{tP}$ differs in sign. However, this sign difference does not significantly affect the theoretical predictions, as the LCDA $\phi_P^t(z, \zeta, \omega^2)$ is twist-suppressed.

We now introduce the $D$-wave two-kaon wave function, as detailed in Refs.~\cite{Zou:2020atb, Li:2020zng}. In the quasi-two-body decays $B^+ \to D_s^+ (K^+K^-)_{D}$, both the initial $B^+$ meson and the final $D_s^+$ meson are pseudoscalars. As a result, the helicity $\lambda = \pm 2$ components of the tensor resonant structure of the $K^+K^-$ pair vanish due to angular momentum conservation. In this case, the $D$-wave wave function behaves similarly to the $P$-wave function, but with a distinct polarization vector derived from the polarization tensor. Thus, the $D$-wave wave function for the two-kaon pair can be decomposed in the same form as in Eq.~\ref{eq:pwave}, involving different light-cone distribution amplitudes (LCDAs):
\begin{widetext}
\begin{align}
    \Phi_D = \sqrt{\frac{2}{3}} \frac{1}{\sqrt{2N_c}} \left[ \slashed{p}_2 \phi_D^0(z, \zeta, \omega^2) + \omega \phi_D^s(z, \zeta, \omega^2) + \frac{\slashed{p}_1 \slashed{p}_2 - \slashed{p}_2 \slashed{p}_1}{\omega (2\zeta - 1)} \phi_D^t(z, \zeta, \omega^2) \right],
\end{align}
where the twist-2 and twist-3 LCDAs are given by:
\begin{align}
    \phi_D^0(z, \zeta, \omega^2) &= \frac{3 F_D^{\parallel}(\omega^2)}{\sqrt{2N_c}} z(1-z) \left[ 3a_D(2z-1) \right] P_2(2\zeta - 1), \\
    \phi_D^s(z, \zeta, \omega^2) &= -\frac{9 F_D^{\perp}(\omega^2)}{4 \sqrt{2N_c}} a_D (1 - 6z + 6z^2) P_2(2\zeta - 1), \\
    \phi_D^t(z, \zeta, \omega^2) &= \frac{9 F_D^{\perp}(\omega^2)}{4 \sqrt{2N_c}} a_D (1 - 6z + 6z^2)(2z - 1) P_2(2\zeta - 1),
\end{align}
\end{widetext}
where the $\zeta$-dependent factor is given by the second-order Legendre polynomial $P_2(2\zeta - 1) = 1 - 6\zeta(1 - \zeta)$. Here, $F_D^{\parallel}(\omega^2)$ and $F_D^{\perp}(\omega^2)$ are the $D$-wave time-like form factors, which will be specified later. The parameter $a_D$ is the Gegenbauer moment, with a value of $0.5 \pm 0.1$, consistent with the result derived from $B \to KKK$ decays in Ref.~\cite{Zou:2020atb}.

As previously mentioned, the RBW line shape serves as an effective model for describing the time-like form factor $F_{S,P,D}(\omega^2)$ associated with narrow-width resonances, reflecting the strong interaction between the two-kaon system and resonant states. The RBW line shape is expressed as \cite{ParticleDataGroup:2022pth}
\begin{align}
    F^L(\omega^2) = \sum_i \frac{c_i m_i^2}{m_i^2 - \omega^2 - i m_i \Gamma_i(\omega^2)},
\end{align}
where the complex coefficients $c_i$ represent the weight and relative phase between different resonance states. The superscript $L$ denotes the spin of the resonant structure. The nominal mass $m_i$ is the mass of the corresponding resonance, while $\Gamma_i(\omega)$ is the mass-dependent width. In the general case of a spin-$L$ resonance, the expression for $\Gamma_i(\omega)$ is given by \cite{BaBar:2012iuj}
\begin{align}
    \Gamma_i(\omega^2) = \Gamma_i^0 \left( \frac{m_i}{\omega} \right) \left( \frac{|\vec{p}|}{|\vec{p}_0|} \right)^{2L+1} X_L^2(r |\vec{p}|),
    \label{eq:gamma}
\end{align}
where $\Gamma_i^0$ denotes the nominal width of the resonance. The values of $m_i$ and $\Gamma_i^0$ for the resonances considered in this work are summarized in Table~\ref{tb:mwidth}. The quantity $|\vec{p}_0|$ is the magnitude of the momentum of one kaon in the $K-K$ pair (denoted $|\vec{p}|$) evaluated at $\omega = m_i$, and $|\vec{p}|$ is the corresponding momentum at arbitrary $\omega$, with the explicit expression
\begin{align}
    |\vec{p}| = \frac{\sqrt{\lambda(\omega^2, m_{h_1}^2, m_{h_2}^2)}}{2\omega},
\end{align}
where the function $\lambda$ is the triangle function, defined as
\begin{align}
    \lambda(x, y, z) = x^2 + y^2 + z^2 - 2xy - 2yz - 2zx.
\end{align}
The Blatt-Weisskopf barrier factor $X_L$ \cite{Back:2017zqt}, which depends on the spin of the resonant structure, is given as
\begin{align}
    &L = 0\;, \; X_0(a) = 1, \\
    &L = 1\;, \; X_1(a) = \sqrt{\frac{1 + a_0^2}{1 + a^2}}, \\
    &L = 2\;, \; X_2(a) = \sqrt{\frac{a_0^4 + 3a_0^2 + 9}{a^4 + 3a^2 + 9}},
\end{align}
where $a_0$ is the value of $a$ evaluated at $\omega = m_i$. The parameter $r$ in Eq.~\ref{eq:gamma} characterizes the effective barrier of the resonance. In experimental analyses \cite{BaBar:2005qms}, $r$ is typically set to a universal value of $4 \, \text{GeV}^{-1} \approx 0.8 \, \text{fm}$ for all resonances, as this choice has a negligible impact on the numerical results.

\begin{table*}[htbp]
\centering
\caption{Parameters describing intermediate states in our framework.}
\label{tab:parameters}
\begin{tabular}{lcccc}
\toprule
\text{Resonance} & \text{Mass (MeV)} & \text{Width (MeV)} & \text{$J^{PC}$} & \text{Model} \\
\hline\hline
$f_0(980)$     & $990$        & $-$          & $0^{++}$ & Flatt\'e  \\
$f_0(1370)$    & $1475$       & $113$        & $0^{++}$ & RBW       \\
$f_0(1500)$    & $1522$       & $108$        & $0^{++}$ & RBW       \\
$\phi(1020)$   & $1019$       & $4.25$       & $1^{--}$ & RBW       \\
$f_2(1270)$    & $1276$       & $187$        & $2^{++}$ & RBW       \\
$f'_2(1525)$   & $1525$       & $73$         & $2^{++}$ & RBW       \\
\hline\hline
\end{tabular}
\label{tb:mwidth}
\end{table*}
\maketitle

Finally, the transverse time-like form factor $F_{P,D}^\perp(\omega^2)$, present in the LCDAs of the $P$-wave and $D$-wave two-meson wave functions, is determined via the approximate relationship
\begin{align}
    \frac{F^\perp}{F^\parallel} \simeq \frac{f_R^T}{f_R},
\end{align}
where $f_R^{(T)}$ denotes the (transverse) decay constant of the resonance. These quantities will be summarized in the following section.

We note that the resonances considered in this work—such as $f_0(1370)$, $f_0(1500)$, $\phi(1020)$, $f_2(1270)$, and $f_2'(1525)$—can be well described by the RBW line shape. In contrast, the simple RBW model fails to describe the $f_0(980)$. This discrepancy arises because the mass of $f_0(980)$ coincides with the $K\bar{K}$ threshold, triggering strong coupled-channel effects and threshold phenomena. As a result, the resonance line shape deviates significantly from the symmetric peak predicted by the RBW formula. To appropriately parameterize this resonance, a multi-channel coupled model, such as the Flatté model \cite{Flatte:1976xu, LHCb:2014ooi}, is required. Thus, we adopt the modified Flatté model \cite{LHCb:2014ooi}, commonly used in experimental amplitude analyses, to describe the $f_0(980)$. The updated expression is given by
\begin{widetext}
\begin{align}
    F(\omega^2) = \frac{m_{f_0(980)}^2}{m_{f_0(980)}^2 - \omega^2 - i m_{f_0(980)} \left( g_{\pi\pi} \rho_{\pi\pi} + g_{KK} \rho_{KK} F_{KK}^2 \right)},
\end{align}
\end{widetext}
where $g_{\pi\pi}$ and $g_{KK}$ are the coupling constants of the $f_0(980)$ to the $\pi\pi$ and $K\bar{K}$ channels, respectively. Their values are typically taken as $g_{\pi\pi} = 0.167 \, \text{GeV}$ and $g_{KK} = 3.47 g_{\pi\pi}$ \cite{LHCb:2014ooi}. Here, $\rho_{\pi\pi}$ and $\rho_{KK}$ are the phase space factors, given by \cite{LHCb:2014ooi}
\begin{align}
    \rho_{\pi\pi} = \sqrt{1 - \frac{4m_{\pi}^2}{\omega^2}}, \quad
    \rho_{KK} = \sqrt{1 - \frac{4m_K^2}{\omega^2}},
\end{align}
with $m_{\pi}$ and $m_K$ denoting the pion and kaon masses, respectively. The factor $F_{KK}$, introduced in Ref.~\cite{Bugg:2008ig} to suppress the $K\bar{K}$ contribution, is factorized as $F_{KK} = e^{-\alpha |\vec{p}_0|^2}$ with $\alpha = 2.0 \, \text{GeV}^{-2}$.

\section{AMPLITUDES and NUMERICAL RESULTS} \label{sec:3}
In Fig.~\ref{fig:placeholder1} and Fig.~\ref{fig:placeholder2}, we present the leading-order Feynman diagrams for the $B^+\to D_s^+(R\to)K^+K^-$ decay in the PQCD framework. Fig.~\ref{fig:placeholder1} shows four emission-type topologies, where the first two correspond to factorizable emission diagrams and the remaining two to nonfactorizable ones. Fig.~\ref{fig:placeholder2} illustrates four annihilation-type topologies, including two factorizable and two nonfactorizable diagrams. Although annihilation contributions are power suppressed relative to emission ones, they provide the dominant strong phase necessary for generating direct $CP$ asymmetry.

Using the factorization formula and the wave functions introduced above, the decay amplitudes can be evaluated systematically. For $B^+\to D_s^+(R\to)K^+K^-$ decays with $S$-wave intermediate resonances, including $f_0(980)$, $f_0(1370)$, and $f_0(1500)$, the contributions from the four classes of topologies—factorizable emission, nonfactorizable emission, factorizable annihilation, and nonfactorizable annihilation—are denoted by $F_{ef,S}$, $M_{enf,S}$, $A_{af,S}$, and $W_{anf,S}$, respectively. For the two factorizable emission diagrams in Fig.~\ref{fig:placeholder1}, the amplitude $F_{ef,S}$ is written as
\begin{widetext}
\begin{align}
F_{ef,S} &= 8\pi C_F m_B^4 f_D 
    \int_0^1 dx_1 dx_2 \int_0^{1/\Lambda} b_1 db_1 \, b_2 db_2 \,
    \phi_B(x_1,b_1) \nonumber \\
&\quad \times \bigg\{ \Big[ (\eta^2-1) \big( \phi_S^{0}(x_2) 
    (r_d^2 (2x_2 + 1) - x_2 - 1) + (2x_2-1)\eta (\phi_S^s(x_2) + \phi_S^t(x_2)) \big) \Big] 
    E_{ef}(t_a) h_{ef}[x_1,x_2,b_1,b_2] \nonumber \\
&\quad + \Big[ \eta (\eta^2 - 1) \big( \phi_S^0(x_2) \eta - 2\phi_S^s(x_2) \big) \Big] 
    E_{ef}(t_b) h_{ef}[x_2,x_1,b_2,b_1] \bigg\},
\end{align}
\end{widetext}
where $b_i\,(i=1,2)$ denotes the variable conjugate to the transverse momentum $k_{iT}$. The Sudakov factor $E_{ef}$ and hard kernel $h_{ef}$ are given in Ref.~\cite{Zou:2015iwa}. Since the emitted $D_s^+$ meson can be factorized and integrated out, its effect is encoded in the decay constant $f_D$, so that $F_{ef,S}$ depends only on the wave functions of the $B$ meson and the $S$-wave $K K$ system. 

By contrast, in the nonfactorizable emission diagrams the hard gluon is attached to both valence quarks of the emitted $D_s^+$, preventing factorization of this meson. Consequently, the amplitude involves all three wave functions and takes the form
\begin{widetext}
\begin{align}
M_{enf,S} &= 16\sqrt{\frac{2}{3}} \pi C_F m_B^4 
    \int_0^1 dx_1 dx_2 dx_3 \int_0^{1/\Lambda} b_1 db_1 \, b_2 db_2 \,
    \phi_B(x_1,b_1)\phi_D(x_3,b_3) \nonumber \\
&\quad \times \bigg\{ \Big\{ -\phi_S^0(x_2)\eta^2 (x_2-2x_3) - \phi_S^0(x_2)x_3   + x_2\eta \big[ \phi_S^s(x_2) - \phi_S^t(x_2) \big] \Big\} 
    E_{enf}(t_a) h_{enf1}[x_1,x_2,b_1,b_2] \nonumber \\
&\quad + \Big[ \eta (\eta^2 - 1) \big( \phi_S^0(x_2) \eta - 2\phi_S^s(x_2) \big) \Big] 
    E_{enf}(t_b) h_{enf2}[x_1,x_2,b_1,b_2] \bigg\},
\end{align}
\end{widetext}
where the corresponding evolution factors and hard functions are also specified in Ref.~\cite{Zou:2015iwa}. In two-body $B$ decays into two light mesons, such as $B\to PP$, $PV$, or $VV$, this class of amplitudes is typically suppressed due to a cancellation between the two nonfactorizable diagrams induced by the relative minus sign of the antiquark propagator. When a $D$ meson is emitted, however, this cancellation becomes ineffective because the large mass difference between the charm quark and a light quark breaks the symmetry between the two diagrams. As a result, for color-suppressed channels the nonfactorizable contributions can even dominate the total decay amplitude~\cite{Zou:2012sy,Zou:2017iau,Zou:2016yhb,Zou:2012sx}.

The diagrams in Fig.~\ref{fig:placeholder2} correspond to annihilation topologies. When the hard gluon attaches to the quarks entering the four-quark operator and subsequently flowing into the final-state mesons, the resulting diagrams are classified as factorizable annihilation diagrams. In this case, the initial $B$ meson can be factorized and integrated out, yielding its decay constant $f_B$. Consequently, the amplitude $A_{af,S}$ depends only on two meson wave functions and can be written as
\begin{widetext}
\begin{align}
A_{af,S}
&= 8\pi C_F m_B^4 f_B 
   \int_0^1 dx_2 dx_3 \int_0^{1/\Lambda} b_2 db_2 \, b_3 db_3 \,
   \phi_D(x_3,b_3) \nonumber \\
&\quad\times \bigg\{ \Big\{
      -\phi_S^0(x_2) \bigl[ x_3 (r_d^2 + 2\eta^2 - 1) + \eta^2 \bigr]
      -\phi_S^s(x_2)\, r_d (x_3+1)\eta
   \Big\} E_{af}(t_a) \; h_{af}[x_2,x_3,b_2,b_3] \nonumber \\
&\quad- \Bigl[
        \phi_S^0(x_2)\bigl( 2r_d + x_2(\eta^2-1) \bigr)
      + \bigl( 1+2r_d x_2 \bigr)\eta\phi_S^s(x_2)
      + 2 r_d (x_2-1) \eta\phi_S^t(x_2)
   \Bigr]E_{af}(t_b) \; h_{af}[x_3,x_2,b_3,b_2] \bigg\}.
\end{align}

If instead the hard gluon is emitted from a quark in the initial $B$ meson, the topology becomes a nonfactorizable annihilation diagram. The corresponding amplitude $W_{anf,S}$ involves all three wave functions and is given by
\begin{align}
W_{anf,S}
&= 16\sqrt{\frac{2}{3}}\pi C_F m_B^4 
   \int_0^1 dx_1 dx_2 dx_3 
   \int_0^{1/\Lambda} b_1 db_1 \, b_2 db_2 \,
   \phi_B(x_1,b_1) \phi_D(x_3,b_2) \nonumber \\
&\quad\times 
   \bigg\{ \Big\{
      -\phi_S^0(x_2) \bigl[ r_d^2 (2x_2-x_3+1) - x_2 + \eta^2 \bigr] - r_d\eta\bigl[\phi_S^s(x_2)(x_2+x_3+2)
                +\phi_S^t(x_2)(x_2-x_3)\bigr]
   \Big\} \nonumber \\
&\qquad\quad\times E_{anf}(t_a) \; h_{anf1}[x_1,x_2,b_1,b_2] \nonumber \\
&\quad- \Bigl[
        \phi_S^0(x_2) x_3
      - r_d\eta\bigl[\phi_S^s(x_2)(x_2+x_3)  + \phi_S^t(x_2) (x_3-x_2)\bigr]
   \Bigr]   E_{anf}(t_b) \; h_{anf2}[x_1,x_2,b_1,b_2] \bigg\}.
\end{align}
\end{widetext}

The $P$-wave and $D$-wave contributions to the $B^+\to D_s^+(R\to)K^+K^-$ decays, associated respectively with the $P$-wave resonance $\phi(1020)$ and the $D$-wave resonances $f_2(1270)$ and $f_2^{\prime}(1525)$, are listed below. The factorizable annihilation amplitude for the $P$-wave case is
\begin{widetext}
\begin{align}
A_{af,P} &= 8\pi C_F m_B^4 f_B 
    \int_0^1 dx_2 dx_3 \int_0^{1/\Lambda} b_2 db_2 \, b_3 db_3 \,
    \phi_D(x_3,b_3) \nonumber \\
&\quad \times \bigg\{ \Big[ \phi_P^0(x_2)x_3(r_d^2+2\eta^2-1) 
    - \phi_P^0(x_2)\eta^2 - 2\phi_P^s(x_2)r_d(x_3+1)\eta \Big]   E_{af}(t_a) h_{af}[x_2,x_3,b_2,b_3] \nonumber \\
&\quad + \Big[ \phi_P^0(x_2)\big(r_d^2(2x_2+1)+x_2(\eta^2-1)\big)   - r_d \eta\big(2\phi_P^s(x_2)x_2+\phi_P^s(x_2)+\phi_P^t(x_2)(2x_2-1)\big) \Big] \nonumber \\
&\quad \times E_{af}(t_b) h_{af}[x_3,x_2,b_3,b_2] \bigg\},
\end{align}
while the nonfactorizable annihilation contribution reads

\begin{align}
W_{anf,P}&= 16\sqrt{\frac{2}{3}} \pi C_F m_B^4 
    \int_0^1 dx_1 dx_2 dx_3 \int_0^{1/\Lambda} b_1 db_1 \, b_2 db_2 \,
    \phi_B(x_1,b_1)\phi_D(x_3,b_2) \nonumber \\
&\quad \times \bigg\{ \Big[ \phi_P^0(x_2)\big(r_d^2(2x_2-x_3+1)-x_2+\eta^2\big) - r_d\eta\big(\phi_P^s(x_2)(x_2+x_3+2)+\phi_P^t(x_2)(x_2-x_3)\big) \Big] \nonumber \\
&\qquad\quad\times E_{anf}(t_a) h_{anf1}[x_1,x_2,b_1,b_2] \nonumber \\
&\quad - \Big[ \phi_P^0(x_2)\eta^2(x_2-2x_3) + \phi_P^0(x_2)x_3  
+ r_d\eta\big(\phi_P^s(x_2)(x_2+x_3)+\phi_P^t(x_2)(x_3-x_2)\big) \Big] \nonumber \\
&\qquad\quad\times E_{anf}(t_b) h_{anf2}[x_1,x_2,b_1,b_2] \bigg\}.
\end{align}

For the $D$-wave sector, the factorizable emission amplitude is
\begin{align}
F_{ef,D} &= 8\sqrt{\frac{2}{3}}\pi C_F m_B^4 f_D 
    \int_0^1 dx_1 dx_2 \int_0^{1/\Lambda} b_1 db_1 \, b_2 db_2 \,
    \phi_B(x_1,b_1) \nonumber \\
&\quad \times \bigg\{ \Big[ (\eta^2-1) \big( \phi_D^{0}(x_2) 
    (r_d^2 (2x_2 + 1) - x_2 - 1) 
    + (2x_2-1)\eta (\phi_D^s(x_2) + \phi_D^t(x_2)) \big) \Big] 
    E_{ef}(t_a) h_{ef}[x_1,x_2,b_1,b_2] \nonumber \\
&\quad + \Big[ \eta (\eta^2 - 1) \big( \phi_D^0(x_2) \eta - 2\phi_D^s(x_2) \big) \Big] 
    E_{ef}(t_b) h_{ef}[x_2,x_1,b_2,b_1] \bigg\},
\end{align}
and the corresponding nonfactorizable emission amplitude is
\begin{align}
M_{enf,D} &= \frac{32}{3}\pi C_F m_B^4 f_D 
    \int_0^1 dx_1 dx_2 dx_3 \int_0^{1/\Lambda} b_1 db_1 \, b_2 db_2 \,
    \phi_B(x_1,b_1) \nonumber \\
&\quad \times \bigg\{ \Big\{ -\phi_D^0(x_2)\eta^2 (x_2-2x_3) - \phi_D^0(x_2)x_3 + x_2\eta \big[ \phi_D^s(x_2) - \phi_D^t(x_2) \big] \Big\} 
    \times E_{enf}(t_a) h_{enf1}[x_1,x_2,b_1,b_2] \nonumber \\
&\quad + \Big[ \eta (\eta^2 - 1) \big( \phi_D^0(x_2) \eta - 2\phi_D^s(x_2) \big) \Big] 
    \times E_{enf}(t_b) h_{enf2}[x_1,x_2,b_1,b_2] \bigg\}.
\end{align}

The factorizable annihilation and nonfactorizable annihilation amplitudes in the $D$-wave case are
\begin{align}
A_{af,D} &= 8\sqrt{\frac{2}{3}}\pi C_F m_B^4 f_B 
   \int_0^1 dx_2 dx_3 \int_0^{1/\Lambda} b_2 db_2 \, b_3 db_3 \,
   \phi_D(x_3,b_3) \nonumber \\
&\quad\times \bigg\{ \Big\{
      -\phi_S^0(x_2) \bigl[ x_3 (r_d^2 + 2\eta^2 - 1) - \eta^2 \bigr]
      - 2\phi_S^s(x_2) r_d (x_3 + 1) \eta
   \Big\} \times E_{af}(t_a) \; h_{af}(x_2,x_3,b_2,b_3) \nonumber \\
&\quad- \Bigl[
        \phi_S^0(x_2) \bigl( r_d^2 (2x_2 - 1) + x_2(\eta^2 - 1) \bigr)
      - \phi_S^s(x_2) \eta \bigl( -2r_d (x_2 + 1) \bigr) \nonumber \\
&\qquad + \phi_S^t(x_2) \eta \bigl( 2r_d (x_2 - 1) \bigr)
   \Bigr] \times E_{af}(t_b) \; h_{af}(x_3,x_2,b_3,b_2) \bigg\},
\end{align}
\begin{align}
W_{anf,D}
&= \frac{32}{3}\pi C_F m_B^4 
   \int_0^1 dx_1 dx_2 dx_3 
   \int_0^{1/\Lambda} b_1 db_1 \, b_2 db_2 \,
   \phi_B(x_1,b_1) \phi_D(x_3,b_2) \nonumber \\
&\quad\times 
   \bigg\{ \Big\{
      -\phi_S^0(x_2) \bigl[ r_d^2 (2x_2-x_3+1) - x_2 + \eta^2 \bigr] - r_d\eta\bigl[\phi_S^s(x_2)(x_2+x_3+2)
                +\phi_S^t(x_2)(x_2-x_3)\bigr]
   \Big\} \nonumber \\
&\qquad\quad\times E_{anf}(t_a) \; h_{anf1}(x_1,x_2,b_1,b_2) \nonumber \\
&\quad- \Bigl[
        \phi_S^0(x_2) \bigl( \eta^2 (x_2 - 2x_3) + x_3 \bigr)   + r_d\eta\bigl[\phi_S^s(x_2)(x_2+x_3) 
                + \phi_S^t(x_2) (x_3-x_2)\bigr]
   \Bigr] \nonumber \\
&\qquad\quad\times E_{anf}(t_b) \; h_{anf2}(x_1,x_2,b_1,b_2) \bigg\}.
\end{align}
\end{widetext}

Using the amplitudes derived above, we obtain the total decay amplitudes, which incorporate the CKM matrix elements and Wilson coefficients. The total $S$-wave amplitudes corresponding to the resonances $f_0(980)$, $f_0(1370)$, and $f_0(1500)$ are summarized as
\begin{widetext}
\begin{align}
    \mathcal{A} (B^{+}\to  D_{s}^{ +}(f_0(980)\to)K^{+} K^{-})&= \mathcal{M}_S^n\sin{\theta}+ \mathcal{M}_S^s\cos{\theta},\\
    \mathcal{A} (B^{+}\to  D_{s}^{ +}(f_0(1370)\to)K^{+} K^{-})&= 0.78\mathcal{M}_S^n+ 0.51\mathcal{M}_S^s,\\
    \mathcal{A} (B^{+}\to  D_{s}^{ +}(f_0(1500)\to)K^{+} K^{-})&= -0.54\mathcal{M}_S^n + 0.84\mathcal{M}_S^s,
\end{align}
\end{widetext}
where the amplitudes $\mathcal{M}_S^n$ and $\mathcal{M}_S^s$ are defined by
\begin{align}
  \mathcal{M}_S^n&= \mathcal{A} (B^{+}\to  D_{s}^{ +}(f_{0} ^{q}\to)K^{+} K^{-})\nonumber\\
  &=\frac{G_{F}}{\sqrt{2} }V_{ub} ^{*} V_{cs }[a_{1}F_{ef,S}+C_{1}M_{enf,S} ],\\
\mathcal{M}_S^s &=\mathcal{A} (B^{+}\to  D_{s}^{ +}(f_{0} ^{s}\to)K^{+} K^{-})\nonumber\\&
=\frac{G_{F}}{\sqrt{2} }V_{ub} ^{*} V_{cs}[a_{1}A_{af,S}+C_{1}W_{anf,S}]. 
\end{align}

The internal structure of the $f_0(980)$ resonance remains unsettled till. Although some experimental evidence favors a tetraquark interpretation, we treat it here as a conventional quark–antiquark state described by the mixing ansatz
\begin{align}
f_0(980)=|n\bar n\rangle \sin\theta + |s\bar s\rangle \cos\theta,
\end{align}
with $n\bar n=(u\bar u+d\bar d)/\sqrt{2}$. The mixing angle $\theta$ is not yet precisely determined. Previous phenomenological analyses of $B\to f_0(980)K^{(*)}$ decays extracted $\theta=17^\circ$, and our recent study of quasi-two-body processes $B\to K/\pi (f_0(980)\to)K^+K^-/\pi^+\pi^-$ showed that this value consistently accommodates current experimental data. We therefore adopt $\theta=17^\circ$ throughout this work.

For the resonances $f_0(1370)$ and $f_0(1500)$, a similar mixing pattern is assumed,
\begin{align}
f_0(1370)&=0.78|n\bar n\rangle+0.51|s\bar s\rangle,\\
f_0(1500)&=-0.54|n\bar n\rangle+0.84|s\bar s\rangle,
\end{align}
where small scalar-glueball components are neglected, since they are beyond the scope of the present analysis.

The total amplitude for the quasi-two-body decay involving the $P$-wave resonance $\phi(1020)$,
\begin{multline}
\mathcal{A}(B^{+}\to D_{s}^{+}(\phi(1020)\to)K^{+}K^{-})\\
=\frac{G_F}{\sqrt{2}} V_{ub}^{*}V_{cs}[a_1 A_{af,P}+C_1 W_{anf,P}],\label{totP}    
\end{multline}
arises purely from annihilation topologies.

Taking into account the mixing between $f_2(1270)$ and $f_2'(1525)$, the corresponding $D$-wave amplitudes are
\begin{eqnarray}
 &&\mathcal{A} (B^{+}\to  D_s^{ +}(f_{2} (1270)\to)K^{+} K^{-})\nonumber\\
 &&\qquad\qquad\qquad\quad=\mathcal{M}_D^n \cos (\theta^{\prime})   +\mathcal{M}_D^s\sin( \theta^{\prime}) ,\\
 &&\mathcal{A} (B^{+}\to  D_s^{ +}(f_{2} (1525)\to)K^{+} K^{-})\nonumber\\
 &&\qquad\qquad\qquad\quad=\mathcal{M}_D^n \sin(\theta^{\prime})-\mathcal{M}_D^s\cos( \theta^{\prime}) ,
 \end{eqnarray}
with $\theta'=7.8^\circ$ and
\begin{align}
\mathcal{M}_D^n&=\mathcal{A} (B^{+}\to  D_{s}^{ +}(f_{2} ^{q}\to)K^{+} K^{-})\nonumber\\
&=\frac{G_{F}}{\sqrt{2} }V_{ub}^{*} V_{cs}[a_{1}F_{ef,D}+C_{1}M_{enf,D}],\\
\mathcal{M}_D^s&=\mathcal{A} (B^{+}\to  D_{s}^{ +}(f_{2} ^{s}\to)K^{+} K^{-})\nonumber\\
&=\frac{G_{F}}{\sqrt{2} }V_{ub}^{*} V_{cs}[a_{1}A_{af,D}+C_{1}W_{anf,D }].
\end{align}

The coefficient $a_1$ denotes the usual combination of Wilson coefficients,
\begin{align}
a_1=C_2+\frac{C_1}{3},
\end{align}
where $C_{1}$ and $C_2$ are the wilson coefficients in the Hamiltonian of Eq.~\eqref{Hmtn}.

After calculating the total decay amplitude, we obtain the differential branching fraction:
\begin{align}
    \frac{d^2\mathcal{B} }{d\zeta d\omega } =\frac{\tau _B\omega \left | \overrightarrow{p_1}  \right |\left | \overrightarrow{p_3}  \right |  }{32\pi^3m_B^3} \left | \mathcal{A}  \right | ^2
\end{align}
where $\tau_B$ is the $B$-meson lifetime. The three-momenta of one kaon and of the bachelor meson $D_s^+$ in the $KK$ center-of-mass frame are
\begin{align}
|\vec p_1|=\frac{\sqrt{\lambda(\omega^2,m_K^2,m_K^2)}}{2\omega},
\qquad
|\vec p_3|=\frac{\sqrt{\lambda(m_B^2,m_D^2,\omega^2)}}{2\omega}.
\end{align}

For numerical evaluation we adopt the following input parameters (masses and decay constants in GeV, lifetime in ps):
\begin{align}
&m_{B^+}=5.280, m_b=4.8, m_c=1.275, m_{K^\pm}=0.494,\nonumber\\\
&  m_{D_s^+}=1.968, f_B=0.19\pm0.02, f_{\phi}=0.215, f_{\phi}^T=0.186,\nonumber\\\
&f_{f_2(1270)}=0.102,
f_{f_2(1270)}^T=0.117,
f_{f_2'(1525)}=0.126, 
\nonumber\\\
&f_{f_2'(1525)}^T=0.065,
\tau_{B^\pm}=1.638 .
\end{align}

\begin{table*}[!htb]
\centering
\caption{PQCD results for the branching ratios of the $S$, $P$ and $D$ wave resonant channels in the $B^+ \to {D^+_{s}} K^+K^-$.}\label{br}
\begin{tabular}{lcc} \hline\hline
{\rm Decay Modes} & & Quasi-two-body\\ \hline
$B^+ \to {D^+_{s}}(f_0(980)\to)K^+K^-$&${\cal B}(10^{-8})$ &$9.90^{+5.61+1.44+0.77}_{-2.92-2.33-0.74}$ \\
$B^+ \to {D^+_{s}}(f_0(1370)\to)K^+K^-$&${\cal B}(10^{-6})$ &$1.04^{+0.86+0.96+0.28}_{-0.70-0.36-0.24}$\\
$B^+ \to {D^+_{s}}(f_0(1500)\to)K^+K^-$ &${\cal B}(10^{-7})$&$1.91^{+0.97+0.54+0.62}_{-0.54-0.42-0.53}$\\
$B^+ \to {D^+_{s}}(\phi(1020)\to)K^+K^-$ &${\cal B}(10^{-8})$&$7.52^{+4.23+1.69+0.94}_{-0.64-2.12-0.58}$\\
$B^+ \to {D^+_{s}}(f_{2}(1270)\to)K^+K^- $&${\cal B}(10^{-7})$&   $6.33^{+4.25+1.14+0.37}_{-3.21-1.21-0.52}$\\
$B^+ \to {D^+_{s}}^0 (f^{\prime}_{2}(1525)\to)K^+K^- $&${\cal B}(10^{-7})$&   $2.13^{+1.49+0.30+0.24}_{-1.34-0.77-0.18}$\\
\hline\hline
\end{tabular}
\end{table*}

Finally, the magnitudes of the isobar coefficients in Eq.~\eqref{ibm} are taken as
\begin{align}
|c_{f_0(1370)}|=0.5, 
|c_{f_0(1500)}|=0.3, 
|c_{f_2(1270)}|=0.3 .
\end{align}
Each coefficient $c_i$ is, in general, complex and encodes the relative phase among different resonant channels. In the present work, however, only their magnitudes are retained, since interference among distinct resonances is not included in our analysis.

Based on the total amplitudes and the input parameters specified above, we evaluate the branching fractions for the quasi-two-body decays $B^+\to D_s^+(R\to)K^+K^-$, where the intermediate resonance $R$ includes the $S$-wave states $f_0(980)$, $f_0(1370)$, and $f_0(1500)$, the $P$-wave state $\phi(1020)$, and the $D$-wave states $f_2(1270)$ and $f_2^{\prime}(1525)$. The numerical results are collected in Table.~\ref{br}. Theoretical predictions in the PQCD framework are subject to several sources of uncertainty. In this analysis we consider three dominant categories. The first, quoted as the leading error in Table.~\ref{br}, arises from variations of hadronic inputs entering the initial- and final-state wave functions. These include the $B$-meson decay constant $f_B$, the shape parameter $\omega_B=0.4\pm0.04~\mathrm{GeV}$ in the $B$-meson LCDA, the decay constant of the $D_s^+$ meson, the parameter $c_D=0.4\pm0.1$ in the $D$-meson wave function, and the Gegenbauer moments appearing in the $S$-, $P$-, and $D$-wave two-kaon distribution amplitudes introduced previously. This source provides the dominant theoretical uncertainty, reflecting the fact that the wave functions constitute the primary nonperturbative inputs of the PQCD formalism. The second category estimates the impact of unknown next-to-leading-order contributions, including radiative and power corrections. We assess this uncertainty by varying the QCD scale $\Lambda_{\mathrm{QCD}}=0.25\pm 0.05~\mathrm{GeV}$ and the typical hard scale $t$ within the interval $0.8t$–$1.2t$. The third source originates from the experimental uncertainties of the CKM matrix elements. The combined effects of these three classes of uncertainties define the total theoretical errors reported in Table.~\ref{br}.

From Table.~\ref{br}, the predicted branching fractions lie in the range $10^{-8}$–$10^{-6}$, which falls within the sensitivity of current and forthcoming measurements at experiments such as LHCb and Belle~II. In particular, the branching fractions for the channels $B^+\to D_s^+(\phi(1020)\to)K^+K^-$ and $B^+\to D_s^+(f_0(980)\to)K^+K^-$ are markedly smaller than those of the remaining modes. For $B^+\to D_s^+(\phi(1020)\to)K^+K^-$, the total amplitude in Eq.~\eqref{totP} shows that this process proceeds purely through annihilation topologies, whose contributions are power suppressed relative to emission diagrams, leading naturally to a reduced rate. By contrast, although $B^+\to D_s^+(f_0(980)\to)K^+K^-$ receives sizable color-allowed emission contributions, its branching fraction remains small because the invariant mass of the final-state kaon pair slightly exceeds the nominal mass of the $f_0(980)$ resonance. In the narrow-width limit this channel would be kinematically forbidden. The finite width of the $f_0(980)$, however, permits the decay through its off-shell tail, but the available phase space is strongly restricted, resulting in an additional suppression of the branching fraction.

\begin{table*}[!htb]
\caption{The branching fractions (in units of $10^{-6}$) for $B^+\to D_s^+R$ decays, extracted from the corresponding quasi-two-body decays based on the Narrow-Width Approximation (NWA), along with the previous PQCD predictions.}\label{narrow}
\label{Tab:Db}
\begin{tabular}{lcc} \hline\hline
{\rm Decay Modes} & NWA & previous PQCD\\ \hline
$B^+ \to {D^+_{s}}f_0(1500)$ &$4.49^{+3.20}_{-1.89}$&$1.99^{+1.53}_{-1.23}$\\
$B^+ \to {D^+_{s}}\phi(1020)$ &$0.15^{+0.09}_{-0.07}$&$0.13^{+0.10}_{-0.08}$\\
$B^+ \to {D^+_{s}}f_{2}(1270)$&$27.52^{+19.19}_{-15.19}$&   $29.9^{+21.6}_{-19.2}$\\
$B^+ \to {D^+_{s}}^0 f^{\prime}_{2}(1525) $&$0.48^{+0.45}_{-0.34}$&   $0.41^{+0.24}_{-0.33}$\\
\hline\hline
\end{tabular}
\end{table*}

To assess the reliability of the two-meson wave-function formalism, we extract the branching fractions of the corresponding two-body decays $B^+\to D_s^+ R$ from the calculated quasi-two-body results using the narrow-width approximation (NWA). Within this approximation, the branching fraction for a quasi-two-body decay $B^+ \to D_s^+(R\to) K^+ K^-$ factorizes into the product of the branching fraction for the underlying two-body transition $B^+\to D_s^+R$ and the decay probability of the intermediate resonance $R\to K^+K^-$. Explicitly,
\begin{multline}
\mathcal{B}[B^+\to D_s^+(R\to K^+K^-)]\\
= \mathcal{B}[B^+\to D_s^+R]
\mathcal{B}[R\to K^+K^-].    
\end{multline}
Using this relation, we combine our predicted quasi-two-body branching fractions with the experimental values for $\mathcal{B}[R\to K^+K^-]$ to determine the corresponding two-body branching ratios $\mathcal{B}[B^+\to D_s^+R]$. Comparison of these extracted results with existing measurements or independent theoretical calculations provides a nontrivial consistency check of the framework. The experimental branching fractions for the relevant resonance decays into $K^+K^-$ are \cite{ParticleDataGroup:2022pth}
\begin{align}
&\mathcal{B}[\phi\to K^+K^-] = (49.9\pm0.5),\\
&\mathcal{B}[f_0(1500)\to K^+K^-] = (4.25\pm0.5),\\
&\mathcal{B}[f_2(1270)\to K^+K^-] = (2.3\pm0.2),\\
&\mathcal{B}[f_2'(1525)\to K^+K^-] = (44.4\pm1.1).
\end{align}
Employing these inputs together with the results in Table.~\ref{br}, we obtain approximate values for the corresponding two-body branching fractions $B^+\to D_s^+R$, which are listed in Table.~\ref{narrow}. For comparison, previously published PQCD predictions \cite{Zou:2016yhb,Zou:2012sx,Zou:2009zza} are also shown. The NWA-extracted results are consistent with earlier PQCD calculations within theoretical uncertainties, providing supporting evidence for the validity of the present treatment.

We do not extract $\mathcal{B}[B^+\to D_s^+f_0(1370)]$, since the experimental value of $\mathcal{B}[f_0(1370)\to K^+K^-]$ is presently unavailable. Likewise, we do not determine $\mathcal{B}[B^+\to D_s^+f_0(980)]$ from the channel $B^+\to D_s^+(f_0(980)\to)K^+K^-$. The NWA is applicable only when the intermediate resonance is narrow and lies sufficiently above the kinematic threshold of its decay products, such that both subprocesses $B\to M_1R$ and $R\to M_2M_3$ can occur on shell. In the present case, the decay $f_0(980)\to K^+K^-$ is kinematically forbidden at the resonance pole. The observed quasi-two-body process instead proceeds through the finite-width tail of the $f_0(980)$, implying that the intermediate state is off shell and cannot be factorized within the NWA. Consequently, the NWA is not applicable to $B^+\to D_s^+(f_0(980)\to)K^+K^-$, and this channel cannot be used to extract $\mathcal{B}[B^+\to D_s^+f_0(980)]$. A determination of this quantity would instead require channels such as $B^+\to D_s^+(f_0(980)\to)\pi\pi$, for which the resonance decay is kinematically allowed.

Finally, we emphasize that no direct $CP$ asymmetries arise for these modes within the SM, since they are mediated exclusively by tree-level operators. Direct $CP$ violation requires interference between amplitudes with different weak phases—typically tree and penguin contributions—which is absent in the present case.

\section{SUMMARY} \label{sec:4}
In this work, we investigate the quasi-two-body decays $B^+\to D_s^+(R\to)K^+K^-$ within the perturbative QCD (PQCD) approach, focusing on contributions from the $S$-wave $K^+K^-$ resonances $f_0(980)$, $f_0(1370)$, and $f_0(1500)$, the $P$-wave resonance $\phi(1020)$, and the $D$-wave resonances $f_2(1270)$ and $f_2(1525)$. The dynamics of the $K^+K^-$ system are described by parameterizing the corresponding timelike form factors $F_{S,P,D}(\omega^2)$ using relativistic Breit--Wigner line shapes and the modified Flatt\'e model, as appropriate. By introducing the relevant $S$-, $P$-, and $D$-wave two-kaon distribution amplitudes to account for the interactions within the kaon pair, we complete the calculation of the decay amplitudes in a consistent factorization framework. As a result, we obtain predictions for the branching fractions of the considered quasi-two-body decay modes, which are found to lie in the range $10^{-8}$ to $10^{-6}$ and are within the reach of current experimental sensitivity. Furthermore, employing the narrow-width approximation, we also extract the branching fractions of the corresponding two-body decays $B^+\to D_s^+R$, providing complementary theoretical predictions for future experimental studies. 

\begin{acknowledgments}
This work is supported in part by the National Science Foundation of China under the Grants Nos. 12375089, 12435004, and 12075086, and the Natural Science Foundation of Shandong province under the Grant No. ZR2022ZD26 and ZR2022MA035
\end{acknowledgments}

\bibliographystyle{bibstyle}
\bibliography{reference}
\end{document}